\documentstyle[preprint,aps,prd]{revtex}

\newcommand{\be}{\begin{equation}}
\newcommand{\ee}{\end{equation}}
\newcommand{\sst}{\scriptscriptstyle}

\newcommand{\bea}{\begin{eqnarray}}
\newcommand{\eea}{\end{eqnarray}}
\def\r0{r_{\sst 0}}

\begin{document}

\title{Scalar Electrodynamics and Primordial Magnetic Fields}

\author{Francisco D.\ Mazzitelli}

\address{{\it
Departamento de F\'\i sica, Facultad de Ciencias Exactas y Naturales\\
Universidad de Buenos Aires- Ciudad Universitaria, Pabell\' on I\\
1428 Buenos Aires, Argentina\\
and\\
Instituto de Astronom\'\i a y F\'\i sica del Espacio\\
Casilla de Correo 67 - Sucursal 28\\
1428 Buenos Aires, Argentina}}

\author{Federico M.\ Spedalieri}

\address{{\it
Departamento de F\'\i sica, Facultad de Ciencias Exactas y Naturales\\
Universidad de Buenos Aires- Ciudad Universitaria, Pabell\' on I\\
1428 Buenos Aires, Argentina\\ }}


\maketitle

\begin{abstract}

A primordial magnetic field  may be generated during an
inflationary period if conformal invariance is broken.
We reexamine and generalize previous results
about the magnetic field produced by couplings
of the form $R^n F_{\mu\nu}F^{\mu\nu}$. We
show that the amplitude of the  magnetic
field depends strongly on $n$. For adequate values of
$n$ the field produced can serve as seed for
galactic magnetic fields.
We also compute the
effective interaction between the electromagnetic field
and the geometry in the context of scalar QED (with and without
classical conformal invariance). In both cases,
the amplitude of the magnetic field is too small
to be of astrophysical interest.

\end{abstract}
\newpage

\section{INTRODUCTION}

Magnetic fields play an important role in a variety of
astrophysical situations. There is
enough evidence for the existence of intragalactic  magnetic fields
\cite{park,zeld},
with an amplitude of $10^{-6}G$ and uniform on a scale of
$10 kpc$. It is not completely clear how these magnetic
fields  were generated. A plausible explanation is that
some kind of dynamo effect \cite{dinamo} could have amplified a
pre-existent magnetic field. But then the question is
about the mechanism that produced this `seed' field.

An attractive suggestion is that it
has a primordial origin and could have been
produced  in the early universe
during an inflationary period \cite{tw,ratra,dolg2}.
Denoting by $r$ the energy density of the magnetic field
relative to the energy density of the cosmic microwave
background radiation, $r={\rho_B\over \rho_{\gamma}}$, a pregalactic
magnetic field caracterized by $r\simeq 10^{-34}$ is needed in
order to explain the present value of $r\simeq 1$.

As pointed out by Turner and Widrow \cite{tw},
it is not possible to produce the required seed field
with the usual
Maxwell Lagrangian. The reason is conformal invariance.
Indeed, in a conformally invariant theory $B$ decreases
as ${1\over a^2}$, where $a$ is the scale factor of the
Robertson--Walker metric. During (exponential) inflation, the
total energy density in the universe is constant,
and the magnetic field energy density is strongly
suppressed, giving $r(\lambda)=10^{-104}({1 Mpc\over \lambda})^4$.

Conformal invariance can be broken in different ways.
{}From a phenomenological point of view, one can
consider \cite{tw} couplings of the form $R A_{\mu}A^{\mu}$.
This type of interaction-terms give rise to the required seed,
but, not being gauge-invariant, are theoretically unappealing.
In string-inspired models, conformal invariance is broken by the
coupling between the electromagnetic
field and the dilaton.
This coupling may produce the seed field \cite{ratra}.
Finally, one can consider gauge-invariant couplings of the
form ${e^2\over m^2}R F_{\mu\nu}F^{\mu\nu}$.
However, the seed field
produced is extremely small \cite{tw}.

In this paper we will reexamine the generation of
primordial magnetic field due to the above mentioned
gauge-invariant couplings. These terms appear due
to quantum effects  when taking into account one loop
corrections
for QED in curved spaces \cite{drum}
(throughout the paper
we will work in the context of scalar QED).  In an expansion
in powers of ${{\cal R}\over m^2}$
(Schwinger DeWitt Expansion, SDWE \cite{sdw})
\footnote {We denote by ${\cal R} $
any component of the Riemann tensor}, one expects
the effective action to contain couplings
the form ${{\cal R}^n\over m^{2n}} F_{\mu\nu}F^{\mu\nu}$.
But as already mentioned in Ref.\cite{tw}, during the
inflationary period one tipically has ${\cal R}\gg m^2$. Therefore,
there is no reason to keep the lowest order contribution $n=1$,
and it is of interest to investigate  more general
couplings.

In the next Section we will compute the primordial magnetic
field produced
by couplings of the form  ${R^n\over m^{2n}} F_{\mu\nu}F^{\mu\nu}$.
For $n=1$, our results are considerably smaller than those
of Ref.\cite{tw}. The discrepancy is due to an overestimation
of the value of $r$ at first horizon crossing. However, we will
show that the amplitude of the magnetic
field depends dramatically
on $n$. In particular, for adequate values of $n$, it is  possible
to generate a  sufficiently large seed field to explain the
present values of the galactic field through a dynamo mechanism.

In this situation, the obvious question is about the
effective Lagrangian in the opposite
regime ${\cal R}\gg m^2$. We address
this issue in Section 3.  Using an
improved version of the SDWE \cite{partoms},
we will show that, in the
leading-logarithm approximation, the effective coupling is
of the form $ F_{\mu\nu}F^{\mu\nu}\, \ln {R\over\mu^2}$.
We will also compute  the amplitude of the magnetic field produced
by this coupling. Unfortunately,
since the logarithm is a slowly varying function,
this amplitude will be extremely small.

In Section 4 we discuss the case
of QED with conformally-invariant quantum fields \cite {dolg2}. Again,
we will find that the
value of $r$ is too small to be the seed of the galactic dynamo.

Throughout the paper we will use units in which $\hbar = c = 1$.

\bigskip

\section{ PRIMORDIAL FIELDS AND THE SCHWINGER DeWITT EXPANSION}

The Lagrangian  for
scalar QED on curved backgrounds
is given by
\begin{equation}
{\cal L}=-{1\over4} F_{\mu\nu}F^{\mu\nu}-D_\mu \phi\, D^\mu \phi^*
+m^2 \phi \phi^* + \xi\, R\,\phi \phi^*
\label{lagqedesc}
\end{equation}
where
$D_\mu = \partial_\mu - ieA_\mu$ is the
covariant derivative for the scalar
field. The theory is conformally invariant only
for massless and conformally coupled  ($\xi = {\scriptstyle {1\over6}}$)
fields.

One can compute an effective Lagrangian for the electromagnetic
field  by integrating out the quantum scalar field.
Using dimensional regularization,
the effective Lagrangian
can be expanded as follows
\cite{birrel}
\begin{equation}
{\cal L}_{eff} =  -{1\over 4}F_{\mu\nu}F^{\mu\nu} +
{1\over2} {1\over (4\pi)^{d\over2}}
\left({m\over \mu}\right)^{d-4}
\sum_{j=0}^\infty a_j(x)\, m^{4-2j}\, \Gamma\, (j-{d\over2})
\label{dessdw}
\end{equation}
where the $a_j(x)$ are the SDW coefficients, and $d$ is
the spacetime-dimension. The first three terms are divergent
in the limit $d\rightarrow 4$, and the poles must be
absorbed into the bare constants of the classical Lagrangian
and into a redefinition of the electromagnetic field.

The first SDW coefficients for a charged scalar field are
given by \cite{barvil}
\begin{eqnarray}
 a_0&=&1\nonumber\\
a_1&=&-\left(\xi - {\scriptstyle {1\over6}}\right)\,R\nonumber\\
 a_2&=&{\scriptstyle{1\over180}}\left(R_{\mu\nu\alpha\beta}\,
       R^{\mu\nu\alpha\beta} -
       R_{\mu\nu}R^{\mu\nu} \right)+ \nonumber \\
    & &+{\scriptstyle{1\over2}}(\xi- {\scriptstyle {1\over6}})^2
       R^2 +
 {\scriptstyle{1\over6}}(\xi- {\scriptstyle {1\over5}})\Box R -
 {e^2 \over 12} F_{\mu\nu}F^{\mu\nu}\nonumber\\
a_3&=&-{\scriptstyle{e^2 \over 45}}F_{\mu\nu ;\rho}F^{\mu\nu ;\rho}
    -{\scriptstyle{e^2 \over 180}}
    F^{\mu\nu}{}{}_{;\nu}F_{\mu\rho}{}{}^{;\rho}
    \nonumber \\
   & &-{\scriptstyle{e^2 \over 30}}\left(\Box F_{\mu\nu}\right)
   F^{\mu\nu}
    +{\scriptstyle{e^3 \over 30}}F_{\mu\nu}F^{\nu\rho}
    F_\rho{}^\mu
    \nonumber \\
   & &+{\scriptstyle{e^2 \over 60}}R_{\mu\nu\rho\sigma}
   F^{\mu\nu}F^{\rho\sigma}
    -{\scriptstyle{e^2 \over 90}}R_{\mu\nu}F^{\mu\rho}F^\nu{}_\rho
    \nonumber \\
   & &+{\scriptstyle{e^2 \over 72}}R F_{\mu\nu}F^{\mu\nu}
    -{\scriptstyle{e^2 \over 12}} \xi R
    F_{\mu\nu}F^{\mu\nu}+\cdots
\label{aj}
\end{eqnarray}
where we omitted the purely
gravitational terms in $a_3$.

After absorbing the poles, the effective Lagrangian
contains the usual Maxwell term plus the finite corrections
\begin{equation}
a_0\,m^4\, \ln{m^2\over\mu^2} +
a_1\,m^2\, \ln{m^2\over\mu^2} + a_2 \ln{m^2\over\mu^2} +
\sum_{k\ge3} {a_k \over (m^2)^{k-2}}\,\Gamma \, (k-2)
\label{partfin}
\end{equation}
{}From the above equations we see that the first non trivial
interactions between the gravitational and electromagnetic
fields are contained in $a_3$ and read
\begin{equation}
R\,F_{\mu\nu}F^{\mu\nu} \ \ ,\ R_{\mu\nu\rho\sigma}\,F^{\mu\nu}
F^{\rho\sigma}\ \ , \   R_{\mu\nu}F^{\mu\rho}F^\nu{}_\rho
\end{equation}
In the same fashion, the SDW coefficient $a_{n+2}$ contains
interactions that can be (schematically)  written  as
${(\nabla)^p \, {\cal R}^{n-p} \over m^{2n}}F^2$.

In what follows, we will analyze the magnetic field produced
by a typical term in the SDWE. For simplicity we
consider the following effective
Lagrangian
\begin{equation}
{\cal L}_{eff}=-{1\over 4}F_{\mu\nu}F^{\mu\nu}
\left ( 1+ b\left ({R\over m^2}\right)^n\right )
\label{rn}
\end{equation}
The classical equations of motion are given by
\begin{equation}
\nabla ^\mu(F_{\mu \nu}(1+\ b\ ({R \over m^2})^n)=0
\end{equation}
In the particular case of a spatially flat Robertson
Walker metric
\begin{equation}
ds^2=a^2(\eta)(-d\eta^2+dx^2+dy^2+dz^2)
\end{equation}
they read
\begin{equation}
\partial ^\mu(F_{\mu \nu}(1+\ b\ ({R \over m^2})^n)=0
\end{equation}
Obviously, if $R$ is constant (exponential inflation),
these equations do not differ from Maxwell equations.
Therefore, non trivial effects appear only for extended inflation.

We will work in the radiation gauge
$A_0=\sum_{i=1}^3 \partial_i A_i =0$
The Fourier expansion of the field is
\begin{equation}
A_j(\eta,\vec x)=\int {d^3k \over (2\pi)^3\,2\omega}
\sum_{\lambda =1}^2 \varepsilon_j^{(\lambda)}(\vec k)\
\left[a^{(\lambda)}(\vec k)\,A_k(\eta)\,e^{i\vec k.\vec x}
+\ h.c.\ \right]
\label{desca1}
\end{equation}
where $\omega= k $, $\ a^{(\lambda)}(\vec k)$ and
$\ a^{\dag\,(\lambda)}(\vec k)$ are
the usual annihilation and creation operators. The vectors
$\ \vec \varepsilon^{\ (\lambda)}(\vec k)$ satisfy
$\ \vec \varepsilon^{\ (\lambda)}(\vec k).\ \vec k =0$.
The classical equation
for the Fourier modes reads\begin{equation}
(\ddot A_k+k^2A_k)(1+\ b\ ({R \over m^2})^n)+\ {b\ n\over(m^2)^n}
R^{n-1} \dot A_k \ \dot R=0
\label{emFc}
\end{equation}
where the dots denote derivatives with respect to the conformal
time $\eta$.

During the inflationary period, $R\gg m^2$ and the quantum correction
dominates over the Maxwell Lagrangian. Indeed, from Einstein equations
we have
\hbox{$R\sim O(\rho_{tot}/m_{pl}^2)$} where $\rho_{tot}$
is the total energy density. Consequently, the term
 \hbox{$b({R/m^2})^nF_{\mu\nu}F^{\mu\nu}$}
dominates over the usual
 \hbox{$F_{\mu\nu}F^{\mu\nu}$} as long as
\hbox{$\rho_{tot}\gg m_{pl}^2\,m^2\,b^{-{1\over n}}$.}
This gives $\rho_{tot}\gg
b^{-{1\over n}}\ (10^8\ GeV)^4$ for
$m=m_e$, the electron mass. In this situation,
Eq. \ref{emFc} reads
\begin{equation}
\ddot A_k+k^2A_k+\ n {\dot R \over R} \dot A_k=0
\label{emFsc}
\end{equation}

Before proceeding, we would like to stress an important point
related to the normalization of  the Fourier modes $A_k$.
As our effective Lagrangian is now
\hbox{$-{b\over 4}({R/m^2})^nF_{\mu\nu}F^{\mu\nu}$},
the canonical commutation relations between the
coordinates $A_i$ and the conjugate momenta
$\pi^i=\dot A^i b\left({R\over m^2}\right )^n$ read
\begin{equation} \left[A_i(\eta,\vec x)\ ,\ \dot A_j(\eta,
\vec{x^\prime}\,)\
b\,\left( {R\over m^2}\right)^n\ \right]=i\left(\delta_{ij}-
{\partial_i\partial_j \over \nabla^2}\right)\ \delta^3(\vec x -
\vec{x^\prime}\,) \label{relconm}
\end{equation}
The usual commutation relations between the creation and
annihilation operators are compatible with the
above commutator only if
\begin{equation}
\dot A_k\ A_k^*-A_k\ \dot A_k^*={2\ \omega\  i \over b \left({R\over
m^2}\right)^n}
\label{norm}
\end{equation}
This will be important in what follows.

Consider the quantity
\begin{equation}
\rho_B ={1\over 8\pi a^4}\  \bigl\langle\  B^2\  \bigr\rangle
\end{equation}
where $\langle\cdots\rangle$
denotes the vacuum expectation value. Although this is not the
magnetic  field energy density associated with the effective
Lagrangian (\ref{rn}),
it becomes the energy density when $\left({R\over m^2}\right)^n \ll 1$.
As a consequence, we can analyze the time evolution of
$\rho_B \over \rho_{tot}$
in order to obtain the value of $r$ at a time when the
Maxwell Lagrangian dominates. For simplicity, we will refer to
$\rho_B$ as the magnetic energy density.

In terms of the
Fourier modes we have
\begin{equation}
\rho_B={1\over 8\,\pi\,a^4}\int {d^3k \over (2\pi)^3\,2\omega}
2\,k^2\ \vert A_k(\eta)\,\vert^2
\end{equation}
The energy density in the $k$th mode of the magnetic field,
defined as $\rho_B(k)=k d\rho_B/dk$, is given by
\begin{equation}
\rho_B(k)={1\over 16\,\pi^3\,a^4}\ k^4\ \vert A_k(\eta)\,\vert^2
\label{rob}
\end{equation}

During extended inflation we can think the Universe as filled
with a perfect fluid having an equation of state
$p=\gamma \rho$, with $-1<\gamma <-1/3$. The scale
factor evolves as $a(t)=\left ({t\over T}\right )^{\alpha}$
where $\alpha = 2/3(1+\gamma) >1$ and $T$ is chosen
so that $a_{today}=1$. This implies that, in conformal
time, $\dot R/R=2/(\alpha -1)\eta$. For this evolution,
the properly normalized solution of Eq.\ref{emFsc} is given by
\begin{equation}
A_k(\eta)=\left(-k\eta\right)^\nu\ \sqrt {\pi\over b}\,
e^{i(\nu+{1\over 2}){\pi\over2}}\,
\left({m^2\ T \over 6\alpha(2\alpha-1)}\right)^{n\over2}\
\left({kT\over {\alpha-1}}\right)^{n\over \alpha-1}\
H_\nu^{(1)}(-k\eta)
\label{sol}
\end{equation}
where  $\nu = {1\over2}- {n \over \alpha -1}$ and $H_\nu^{(1)}(x)$
is a Hankel function.
In order to select a unique solution we have  also
imposed that $A_k(\eta)\sim e^{-i k \eta}$ as
$\eta\rightarrow -\infty$. This is the natural
choice for the vacuum state in the $in$ region.

First horizon crossing takes place when $-k\eta=2\pi\alpha/
(\alpha -1)$.  Inserting this value of $\eta$ in
Eqns.\ref{rob} and \ref{sol} one finds the value of $\rho_B$ at that
time. The value of $\rho_{tot}(\eta)$ can be obtained
from Einstein equations at first horizon crossing. It is given by
\begin{equation}
{\rho_{tot}(\lambda) }={m_{pl}^2 \over 6\pi}\
\left({4\pi \over 3}\right)^{{2\over \alpha -1}}\
\left({3\alpha \over 2}\right)^{{2\alpha \over \alpha -1}}\
\left({\lambda \over 2\pi T}\right)^{{2\over \alpha-1}}\
T^{-2}
\label{condin}
\end{equation}
The value of $r$ at first horizon crossing, denoted
by $\r0$, is therefore
given by $\r0= {\rho_B \over \rho_{tot}}\bigg\vert_{1^{st} crossing}.$
After first horizon crossing the quantum fluctuations are assumed to become
classical perturbations. For $k\eta\gg 1$
the term $k^2A_k$ in Eq.\ref{emFsc} can be neglected. The time evolution
of the Fourier modes is approximately given by
\begin{equation}
\ddot A_k+\ {n\ s \over \eta} \dot A_k =0
\label{s}
\end{equation}
where $s=-6(1+\gamma)/(1+3\gamma)$.

{}From here, the analysis follows closely   that of Ref.\cite{tw}.
Eq.\ref{s} admits a constant solution, which gives an
uninteresting $\rho_B\sim a^{-4}$. It also has the
solution
\begin{equation}
A_k\sim \eta^{[(6n+1)+(6n+3)\gamma]\over
1+3\gamma}\,\,\sim\left(a \right)^
{{1\over2}[(6n+1)+(6n+3)\gamma]}
\label{soluc}
\end{equation}
The solution
grows with time for $\gamma >-{6n+1\over 6n+3}$. In this
situation, and as long as $R\gg m^2$, $r$ evolves
like $\rho_{tot}^{-{2n+ 2\gamma (n+1)\over 1+\gamma}}$.
When $R\simeq m^2$, or when conductivity effects during reheating
\cite{tw}
become dominant, $\rho_B\sim a^{-4}$ and
$r\ \propto \ \rho_{tot}^{(1-3\gamma) \over 3(1+\gamma)}$.
Assuming a very rapid reheating, the Universe enters the
radiation dominated period, $\rho_{tot}$ equals $\rho_{\gamma}$,
and $r$ becomes a constant. In order to calculate this final
value of $r$ we have to distinguish between two possibilities.
The first one corresponds  to the case in which the growing
solution disappears before the plasma effects become dominant.
This means that $M^4 < b^{-{1\over n}}\, m_{pl}^2 \, m^2$,
where  $M^4$
is the total energy density at the end of the inflationary
period, and
$b^{-{1\over n}}\, m_{pl}^2 \, m^2$ is the total energy density at
the instant in which
the $R^n\,F^2$ term ceases to dominate over the usual Maxwell term. In this
case, and using the  relations betwen $r$ and $\rho_{tot}$ stated above,
we get that the value of $r$ at the beginning of the radiation dominated
period is given by
\begin{equation}
r=\r0\ \left({\rho_{tot}(k)  \over b^{-{1\over n}}\ m_{pl}^2\ m^2}
\right)^{2n+2(n+1)\gamma \over 1+\gamma}\
\left({M^4 \over b^{-{1\over n}}\ m_{pl}^2\ m^2} \right)^{1-3\gamma
\over 3(1+\gamma)}
\label{rcomp1}
\end{equation}
The other possibility is that the conductivity becomes important
before the growing solution disappears, i.e., that $M^4 > b^{-{1\over n}}
\, m_{pl}^2 \, m^2$.
Then, $r\propto \rho_{tot}^{-{2n+ 2\gamma (n+1)\over 1+\gamma}}$
during all the
period that goes from first horizon crossing to the end of inflation, and
therefore
the final value of $r$ reads now
\begin{equation}
r=\r0\ \left({\rho_{tot}(k) \over M^4} \right)^{2n+2(n+1)\gamma
\over 1+\gamma}
\label{rcomp2}
\end{equation}
For $b=O(1)$ and $m=m_e$, we found that there is a value of $\gamma$
between $-0.7$ and $-0.6$ that separates the two cases.
The results for different values of $\gamma$, $\lambda = 1\ Mpc$ and  $m=m_e$
are summarized in Table I.

\vskip 1cm
\begin{center}
\begin{tabular}{|c|c|c|c|c|c|}
\hline
\multicolumn{6}{|c|}{Table I} \\\hline
\rule[-1ex]{0pt}{4ex}
$\gamma$ & \multicolumn{5}{|c|}{$r \vert_{1\ Mpc}$} \\\hline
\rule[-1ex]{0pt}{4ex}
 & $n=1$ & $n=2$ & $n=3$ & $n=4$ & $n=5$  \\
\hline
\rule[-1ex]{0pt}{5ex}
$ -0.5$ & $10^{-84} $ & $10^{-67}$
& $10^{-51}$ & $10^{-35}$ &
$10^{-19}$
\rule[-1ex]{0pt}{4ex} \\
 $ -0.6$ & $10^{-92} $ & $10^{-63}$ &
 $10^{-37}$ & $10^{-10}$ & $10^{17}$
\rule[-1ex]{0pt}{4ex} \\
$ -0.7$ & $10^{-116} $ & $10^{-96}$ &
$10^{-77}$ & $10^{-56}$  & $10^{-36}$
\rule[-1ex]{0pt}{4ex} \\
$ -0.8$ & $(10^{-141} )$ & $10^{-146}$ &
$10^{-147}$ & $10^{-148}$  & $10^{-149}$
\rule[-1ex]{0pt}{4ex} \\
$ -0.9$ & $(10^{-184} )$ & $(10^{-222})$ &
$10^{-253}$ & $10^{-269}$  & $10^{-285}$
\rule[-1ex]{0pt}{4ex} \\ \hline
\end{tabular}
\end{center}
\vskip 0.5cm
\hskip 0.5cm Table I: $r\vert_{1\,Mpc}$ as a function of
$\gamma$ and $n$. The numbers between brackets  correspond to
situations where there is no growing solution. The value
of $r$ for $n=5$ and $\gamma =-0.6$ cannot be trusted
because it does not satisfy
the condition $r\ll 1$.

\vskip 0.5cm

These results depends strongly on $n$. For
$\gamma >-.6$, we are in the case in which the
growing solution disappears before the conductivity becomes
dominant. We can see from Table I and from Eq.\ref{rcomp1}
that the value of $r$ increases exponentially with $n$ (for $n$
high enough, the resultant field can serve
to seed the galactic dynamo). For $\gamma <-.7$,
plasma effects dominates before the growing solution
disappears, and therefore the period of time during
which the amplification mechanism is operative, is
less than the corresponding one in the opposite case. Then,
the value of $r$ as given by Eq.\ref{rcomp2}, is a
compromise between the amplification and the time during
which it takes place. In Table I we can see this behavior:
for $\gamma =-.7$ we still have an exponential growth
of $r$, while for $\gamma <-.8$ the period of
amplification is not enough and $r$ now decreases exponentially
with $n$.

It may seem strange that we obtain an exponential
decrease in certain cases while we can see from
Eqs.\ref{rcomp1} and \ref{rcomp2} that the amplification
factor grows exponentially with $n$. The cause of
this behavior is in the dependence of the initial
condition  $\r0$ on $n$ (see Table II).

\vskip 1cm

\begin{center}
\begin{tabular}{|c|c|c|c|c|c|}
\hline
\multicolumn{6}{|c|}{Table II} \\\hline
\rule[-1ex]{0pt}{4ex}
$\gamma$ & \multicolumn{5}{|c|}{$\r0 \vert_{1\ Mpc}$}\\\hline
\rule[-1ex]{0pt}{4ex}
 & $n=1$ & $n=2$ & $n=3$ & $n=4$ & $n=5$  \\
\hline
\rule[-1ex]{0pt}{5ex}
$ -0.5$ & $10^{-41} $ & $10^{-58}$ &
$10^{-76}$ & $10^{-94}$ & $10^{-112}$
\rule[-1ex]{0pt}{4ex} \\
  $-0.6$ & $10^{-43} $ & $10^{-71}$ &
  $10^{-100}$ & $10^{-128}$ & $10^{-157}$
\rule[-1ex]{0pt}{4ex} \\
$ -0.7$ & $10^{-44} $ & $10^{-78}$ &
$10^{-112}$ & $10^{-146}$  & $10^{-180}$
\rule[-1ex]{0pt}{4ex} \\
$ -0.8$ & $10^{-43} $ & $10^{-78}$ &
$10^{-113}$ & $10^{-148}$  & $10^{-183}$
\rule[-1ex]{0pt}{4ex} \\
$ -0.9$ & $10^{-45} $ & $10^{-83}$ &
$10^{-121}$ & $10^{-159}$  & $10^{-197}$
\rule[-1ex]{0pt}{4ex} \\ \hline
\end{tabular}
\end{center}

\vskip 0.5cm
\hskip 0.5cm {Table II : \  $\r0\vert_{1\ Mpc}$ for the
electromagnetic field, for different values of $\gamma$ and $n$}

\vskip 0.5cm

\noindent As we see, $\r0$ decreases exponentially with $n$, so the
behavior of the final value of $r$ with $n$ is a compromise between
the decrease of the initial condition and the growth of
the amplification factor, both of which depend on $\gamma$.

At this point, we would like to compare
our results for $n=1$, with those of Ref.\cite{tw}.
Our values for $r$ are smaller by
several orders of magnitude
(compare our $r(\gamma=-.6)\sim 10^{-92}$ with the
correspondent result $r\sim 10^{-68}$ in Ref.\cite{tw}).
It is easy to find the origin of the discrepancy.
The calculation in Ref.\cite{tw} begins after
first horizon crossing and assumes
that the initial value of $\rho_B$ is
$H^4$. This is the value of the energy density
for a massless minimally coupled scalar field. Using
this value the resulting $\r0$ varies between $10^{-33}$ and $10^{-12}$
(for $-0.9 \leq \gamma < -0.5$).
Here we   computed $\rho_B\vert_{1st\, crossing}$
from first principles, and found much smaller
values (see Table II). The physical
origin  of the discrepancy is the normalization
condition Eq.\ref{norm}.

Up to here, we have made our calculations using $m=m_e$.
Now we turn to study the dependence on $m$ of our results.
The mass $m$ enters the calculation at two different points. First,
it determines when the $\left({R \over m^2 }\right)^n\ F^2$
coupling ceases to
dominate over the usual Maxwell term, and hence it
contributes to determine the amount of amplification
that will take place. Second, it appears in the initial
condition in such a way that
$\r0 \propto m^{2n}$. We can see that raising the value
of $m$ improves the initial condition $\r0$. On the other hand,
it makes shorter the amplification period, and the final result
is a compromise between these two effects. We have analysed the mass
dependence taking these facts into account. We have found that
for $M^2 < b^{-{1\over n}}\,m_{pl}\,m$, the results decrease
with increasing mass, while for $M^2 > b^{-{1\over n}}\,m_{pl}\,m$
the results increase with mass. In the first case the
shortening of the amplification period preponderates over the
increasing of the initial condition, making the result decreasing with
increasing mass. In the other case, the duration of the
amplification period depends only on the value of $M$ and no
longer on $m$. This means that raising the value of the mass
improves the initial condition but does not affect the amplification,
giving an increasing result with increasing mass. However,
in this case the value of $m$ cannot be
raised arbitrarily, because it must obey the relation
$M^2 > b^{-{1\over n}}\,m_{pl}\,m$.

To summarize, although
the value of
$\r0$ is strongly suppressed with respect to
that of a minimally coupled scalar field,
and although the suppression increases with $n$,
there are values of $\gamma$ such that, after first horizon crossing,
a large amplification
takes place, and produces final values of
$r$ large enough to serve as the seed magnetic field.

\bigskip

\section{IMPROVING THE SDWE -- KILLING THE MAGNETIC FIELD}

{}From a phenomenological point of view, the couplings
discussed in the previous section may help to solve
the problem of the generation of  a primordial field
whitout breaking gauge invariance.
Do they have  a
theoretical motivation? In order to answer this question,
we need to compute the effective action for the electromagnetic
field in the limit ${\cal R}\gg m^2$.
To do this,
we will use the improved version of
the SDWE  developed  by Parker and Toms in Ref.\cite{partoms}.

A partial ressumation of the SDWE can be achieved by
doing the expansion in inverse powers of
$\tilde{m}^2=m^2 + (\xi -{1\over 6})R$ instead of inverse
powers of $m^2$. In this case, as conjectured in Ref.\cite
{partoms} and proved in Ref.\cite{jack}, the {\it new}
SDW coefficients (denoted here by $b_j$) do not contain
powers of the scalar curvature $R$. The first $b_j$
are given by
\begin{eqnarray}
b_0 &=& 1\nonumber\\
b_1 & = & 0\nonumber\\
b_2 &=& {\scriptstyle {1\over180}}\left(R_{\mu\nu\alpha\beta}\,
R^{\mu\nu\alpha\beta} -
R_{\mu\nu} R^{\mu\nu}  \right)+
{\scriptstyle{1\over6}}(\xi-{\scriptstyle{1\over5}})\Box R-
{e^2\over 12} F_{\mu\nu}F^{\mu\nu} \nonumber\\
b_3&=&-{\scriptstyle{e^2\over 45}}F_{\mu\nu ;\rho}F^{\mu\nu ;\rho}
    -{\scriptstyle{e^2 \over 180}}F^{\mu\nu}{}{}_{;\nu}
    F_{\mu\rho}{}{}^{;\rho}-
    \nonumber \\
   & &-{\scriptstyle{e^2 \over 30}}\left(\Box
   F_{\mu\nu}\right) F^{\mu\nu}
    +{\scriptstyle{e^3 \over 30}}F_{\mu\nu}F^{\nu\rho}F_\rho{}^\mu +
    \nonumber \\
   & &+{\scriptstyle{e^2 \over 60}}R_{\mu\nu\rho\sigma}
   F^{\mu\nu}F^{\rho\sigma}
    -{\scriptstyle{e^2 \over 90}}R_{\mu\nu}
    F^{\mu\rho}F^\nu{}_\rho+\cdots
\label{b3}
\end{eqnarray}
where we  did not include the purely gravitational
terms in $b_3$.  Note that $b_1$ and $b_2$ do not
contain the terms proportional to $R$ and $R^2$
that are present in $a_1$ and $a_2$ (see Eq.\ref{aj}).
The same thing happens for all $b_j$.

The improved expansion can be obtained from the usual
SDWE (Eq.\ref{dessdw})  substituting $m^2$ by $\tilde{m}^2$ and
the coefficients $a_j$ by the $b_j$.
In the limit $(\xi-{\scriptstyle {1\over6}})R\gg m^2$,
the coefficient $b_2$ induces a coupling of the form
\begin{equation}
{1\over4}\left({e^2 \over 96\pi^2}\right)F_{\mu\nu}F^{\mu\nu}
\ln\left({R\over \mu^2}\right)
\end{equation}
Under the assumption
$\ln {R\over\mu^2}\gg 1$, this term  dominates and the effective
lagrangian
reduces to
\begin{equation}
{\cal L}_{eff}=-{1\over4}F_{\mu\nu}F^{\mu\nu}
\left(1-{e^2 \over 96\pi^2}\ln {R\over \mu^2}\right)+...
\label{lagef}
\end{equation}
where $\mu$ is an arbitrary scale.
It is worth noting that this  result  is also
valid beyond one-loop, in the  leading-logarithm approximation.
A similar effective lagrangian
has been proposed for non-abelian gauge theories in Ref.\cite
{calzetta}.

There is a simple physical interpretation of the effective Lagrangian
given by Eq. \ref{lagef}. If we couple ${\cal L}_{eff} $
to an external current, ${\cal L}_{eff} \rightarrow {\cal L}_{eff} + e
J_{\mu}A^{\mu}$, after a rescaling of $A^{\mu}$ we obtain
the usual Maxwell Lagrangian with a running
(curvature-dependent) electric charge
\begin{equation}
e^2(R)={e^2(\mu^2)\over 1-{e^2 \over 96\pi^2}\ln {R\over\mu^2}}
\end{equation}
This is  the scale dependence dictated by the renormalization group.
This interpretation also makes explicit the fact that ${\cal L}_{eff}$
is valid only under the assumption  ${e^2 \over 96\pi^2}\ln {R\over\mu^2}
\ll 1$. Otherwise, an analysis of strongly coupled QED is needed
(strictly speaking, for such large values of the curvature
one should replace QED by a GUT).

We will perform the same calculation we did in the previous Section, that is,
calculate the value of $r$ for a scale of $1\, Mpc$.
Typically, this scale  crosses outside the horizon not long after curvature
falls below the GUT scale ($10^{28} \, GeV^2$). Thus, we cannot
use the effective lagrangian (\ref{lagef}) to calculate the
initial condition, so we will consider $\r0$ as given and study its
evolution after first horizon crossing.

The equation for Fourier modes is
\begin{equation}
\ddot A_k+k^2A_k-{e^2 \over 96 \pi^2}\,
{\dot R \over R} {\dot A_k \over \left(1-{e^2
\over 96 \pi^2} \ln {R\over \mu^2}\right)}=0
\label{mf}
\end{equation}
We will take $\mu^2 = R_{RH}$, the value of the scalar curvature
at the end of reheating, and we will consider that $e$, the
electric charge when $R=R_{RH}$, is of the same order of
magnitude of its value today. Then we can assure that,
during most of inflation,  $\ln {R\over R_{RH}}
\gg 1$ and ${e^2 \over 96 \pi^2} \ln {R\over R_{RH}} \ll 1$,
the conditions needed for the validity of the leading logarithm
 approximation. In this situation,
for modes outside the horizon, we can approximate
Eq. \ref{mf} by
\begin{equation}
\ddot A_k-{e^2 \over 96 \pi^2}\, {\dot R \over R}\, \dot A_k =0
\label{mf2}
\end{equation}
This equation has the same form of Eq. \ref{s}, if we make the
identification $n\leftrightarrow -{e^2\over 96 \pi^2}\equiv x$.
The solutions are the same as those of Eq. \ref{s}, that is, $A_k = constant$,
which gives $\rho_B \propto a^{-4}$, and
\begin{equation}
A_k\sim \left(a \right)^{{1\over2}[(6x+1)+(6x+3)\gamma]}
\label{sol2}
\end{equation}
Using that $x\sim 10^{-4}$ we can write Eq. \ref{sol2} as
\begin{equation}
A_k\sim \left(a \right)^{{1\over2}[1+3\gamma]}
\end{equation}
Since $1+3\gamma < 0$, this means that,
for this solution, $\rho_B$ decreases more
rapidly than $a^{-4}$. So we can consider only the contribution
of the constant solution and conclude that for the effective
lagrangian given by Eq. \ref{lagef}, $\rho_B$
decreases as $a^{-4}$, and then $r \propto  \rho_{tot}^{(1-3\gamma)
\over 3(1+\gamma)}$.

The final value of $r$ will be
\begin{equation}
r= \r0 \left({M^4 \over \rho_{tot} (k)}\right)^{(1-3\gamma) \over 3(1+\gamma)}
\end{equation}
For $\gamma < -0.5$, the amplification factor that accompanies $\r0$ is
smaller than $10^{-49}$. Since the initial condition is tipically
much less than $1$,
the value of $r$ is in all cases  very far
from the required value $10^{-34}$.

To conclude, although the higher order terms in the SDWE
${\cal R}^n F^2$
generate large amplitudes for the magnetic field, once
these terms are ressumed they generate only a logarithmic
interaction $\ln R\ F^2$. This interaction   produces
a discouragingly small seed field.

\section{TRACE ANOMALY AND PRIMORDIAL FIELDS}

We end the paper with a comment about the conformally
invariant case $m^2=0$, $\xi =1/6$. Of course in this
situation neither the SDWE nor its improved version are
useful to compute the effective Lagrangian. However,
it is easy to find a closed expression for it.

It is
well known that conformal invariance is broken by
quantum effects. The (anomalous) trace of the energy momentum
tensor   is given by the second SDW coefficient \cite{birrel}
\begin{equation}
T_{\mu}^{\,\mu}=- {a_2\over 16\pi^2}={e^2\over 192\pi^2}
F_{\mu\nu}F^{\mu\nu} + ...
\label{trace}
\end{equation}
where the dots denote purely gravitational
terms proportional to ${\cal R}^2$ and $\nabla\nabla\cal R$.

In a Robertson Walker space-time, the above trace anomaly
can be derived from the effective
Lagrangian \cite{dolg1}
\begin{equation}
{\cal L}_{eff}=-{1\over 4}F_{\mu\nu}F^{\mu\nu}\left(
1+\kappa\ln {a\over a_0}\right ) + ....
\label{conf}
\end{equation}
where $\kappa={e^2\over 48\pi^2}$ and $a_0$ is a reference value
for the scale factor
(for example the value of $a$ after inflation).
Here the dots denote non-local terms independent of the scale
factor.
\footnote {There is a sign difference between the
logarithmic term of the effective lagrangian in Ref. \cite{dolg1}
and that of Ref. \cite{dolg2}.
We agree with the result of Ref. \cite{dolg1}.}

As before, Eq.\ref{conf} can be interpreted as the usual
Maxwell Lagrangian with a scale dependent electric charge
\begin{equation}
e^2(a)={e^2(a_0)\over (1+\kappa\ln
{a\over a_0})}
\label{running}
\end{equation}
The effective Lagrangian is valid as long as
$\kappa\ln
{a\over a_0}\ll 1$.

Under this assumption, one can estimate
the magnetic field generated
by the conformal anomaly. The resulting value for $r$
is again extremely small, of order $10^{-104}$
for $\lambda\sim 1Mpc$.

It has been argued \cite{dolg2}  that this could be an
efficient mechanism if a large number of massless
($m\ll H$) fields are present during inflation,
 so that $\kappa\rightarrow N\kappa\sim O(1)$.
Using the approximation
\begin{equation}
{\cal L}_{eff}\simeq -{1\over 4}F_{\mu\nu}F^
{\mu\nu}\left({a\over a_0}\right )^{N\kappa}
\label{leffaprox}
\end{equation}
it is possible
to show \cite{dolg2} that the previous result for $r$ is
multiplied by a
huge amplification
factor   $A= \left ({H\over k}\right )^{N \kappa}\sim 10^{50 N\kappa}$
for $H=10^{12} GeV$ and $\lambda =1 Mpc$.

However, during that period $\ln{a\over a_0}$ varies
between $-60$ and $0$ and the approximation (\ref{leffaprox})
breaks down for $N\kappa\sim 1$.
In this situation, even
the effective Lagrangian (\ref{conf}) is
inadequate, since at some point $1+N\kappa\ln{a\over a_0}$
vanishes and one reaches the Landau singularity.
Therefore, the conclusion is that, if $N\kappa$ is
sufficiently small in order to assure that $N\kappa\ln{a\over a_0}\ll 1$
during the period of interest, the amplification factor
is very small, $A\simeq 1+N\kappa\ln{H\over k}$.
The amplitude of the magnetic field generated
is  too small to be of
astrophysical interest.
On the other hand, if  $N\kappa\sim O(1)$, the final result
for $r$ would depend on the physics of QED at strong coupling.

\section{Acknowledgments}

We would like to thank D.D. Harari, M. Zaldarriaga and
E. Calzetta
for useful conversations.
This research was supported by Universidad de Buenos Aires,
Consejo Nacional de Investigaciones Cient\'\i ficas y T\' ecnicas
and by Fundaci\' on Antorchas.

\end{document}